\documentclass[preprint,aps,prl,showpacs,superscriptaddress]{revtex4-1}
\selectfont

\usepackage{braket}
\usepackage{bm}
\usepackage{graphicx}
\graphicspath{{pict/}{}}

\usepackage{float}
\usepackage{amsmath}
\usepackage{color}

\usepackage{hyperref}
\hypersetup{pdfstartview={FitH},pdfpagemode={UseNone},
            colorlinks,linkcolor=blue, citecolor=blue, urlcolor=blue,
            bookmarks=true, bookmarksopen=true, pdfnewwindow=true}

\usepackage{natbib}
\newcommand\hrefBibPDF[3][]{}

\begin{document}
\date{\today}
	\title{Scalable on-chip quantum state tomography}

\author{James Titchener}
\email{james.titchener@anu.edu.au}
\affiliation{Nonlinear Physics Centre, Research School of Physics and Engineering, Australian National University, Canberra, ACT 2601, Australia}
\affiliation{Institute of Applied Physics, Abbe Center of Photonics, Friedrich-Schiller-Universität Jena, Max-Wien-Platz 1, Jena 07743, Germany}

\author{Markus Gr{\"a}fe}
\author{Ren{\'e} Heilmann}
\affiliation{Institute of Applied Physics, Abbe Center of Photonics, Friedrich-Schiller-Universität Jena, Max-Wien-Platz 1, Jena 07743, Germany}

\author{Alexander~S.~Solntsev}
\affiliation{Nonlinear Physics Centre, Research School of Physics and Engineering, Australian National University, Canberra, ACT 2601, Australia}

\author{Alexander Szameit}
\affiliation{Institute for Physics, University of Rostock, Universit¨atsplatz 3, 18055 Rostock, Germany}
\affiliation{Institute of Applied Physics, Abbe Center of Photonics, Friedrich-Schiller-Universität Jena, Max-Wien-Platz 1, Jena 07743, Germany}

\author{Andrey A. Sukhorukov}
\affiliation{Nonlinear Physics Centre, Research School of Physics and Engineering, Australian National University, Canberra, ACT 2601, Australia}

\begin{abstract}
\textbf{
		Quantum information systems are on a path to vastly exceed the complexity of any classical device.
	The number of entangled qubits in quantum devices is rapidly increasing \cite{Barends:2014-500:NAT, Wang:2016-210502:PRL, Monz:2011-130506:PRL} and the information required to fully describe these systems scales exponentially with qubit number \cite{Nielsen:2011:QuantumComputation}.
		This scaling is the key benefit of quantum systems, however it also presents a severe challenge. To characterize such systems typically requires an exponentially long sequence of different measurements \cite{Altepeter:2005-105:RAR}, becoming highly resource demanding for large numbers of qubits \cite{Banaszek:2013-125020:NJP}.
		Here we propose a novel and scalable method to characterize quantum systems, where
	the complexity of the measurement process only scales linearly with the number of qubits.
		We experimentally demonstrate an integrated photonic chip capable of measuring
	two- and three-photon quantum states with reconstruction fidelity of $99.67\%$.
	}
\end{abstract}

\maketitle

The standard way to characterize a quantum system is known as quantum state tomography \cite{Vogel:1989-2847:PRA, Altepeter:2005-105:RAR}. It involves measuring expectation values of a complete set of observables and using these to reconstruct the system's density matrix \cite{James:2001-52312:PRA, Haffner:2005-643:NAT, Leibfried:2005-639:NAT, Lvovsky:2009-299:RMP, Bayraktar:2016-20105:PRA, Shadbolt:2012-45:NPHOT}.
To characterize an $N$-qubit state, $2^{2N}$ different observables are measured \cite{James:2001-52312:PRA}, thus the measurement apparatus must be reconfigured exponentially many times, which is impractical for large states.
Furthermore many of the expectation values measured will be vanishingly small, and thus contribute little useful information. Finally, even if all the measurements can be completed, the task of reconstructing the density matrix from measurement data becomes computationally challenging
%intractable 
%\textcolor{red}{reference what means intractable}
for high qubit-number states \cite{Banaszek:2013-125020:NJP}. 

New approaches to quantum state tomography are 
being developed in an effort to 
increase its practicality and efficiency. 
%It was shown 
Some approaches seek to avoid unnecessary measurements by assuming that the system is in particular low rank states, such as sparse states \cite{Gross:2010-150401:PRL,Oren:2016-226:OPT} or low dimensional matrix product states \cite{Cramer:2010-149:NCOM}.
Alternatively tomography can be `self-guided', where real-time feedback of measurement results  guides the next choice of the measurement basis \cite{Ferrie:2014-190404:PRL, Chapman:2016-40402:PRL}, helping to avoid taking  measurements that have limited utility for analyzing the state.
It has been shown that tomography procedures involving some global quantum measurements have  increased error robustness relative to using only local qubit measurements, and thus can be completed in less time \cite{Miranowicz:2014-62123:PRA}.
The computational burden of inverting large data sets to find the density matrix is reduced with simple real-time optimization algorithms in self guided tomography, or can be completely avoided with systems for direct projection of density matrix parameters \cite{Lundeen:2011-188:NAT,Bolduc:2016-10439:NCOM, Thekkadath:2016-120401:PRL}. 
However all these approaches rely on a common measurement paradigm, whereby different characteristics of a system are measured sequentially, thus they become exponentially complex to implement as the number of parameters in the density matrix scales exponentially with qubit number.

%\begin{figure*}[bt]
\begin{figure*}
	\centering
  	\includegraphics[width=\textwidth]{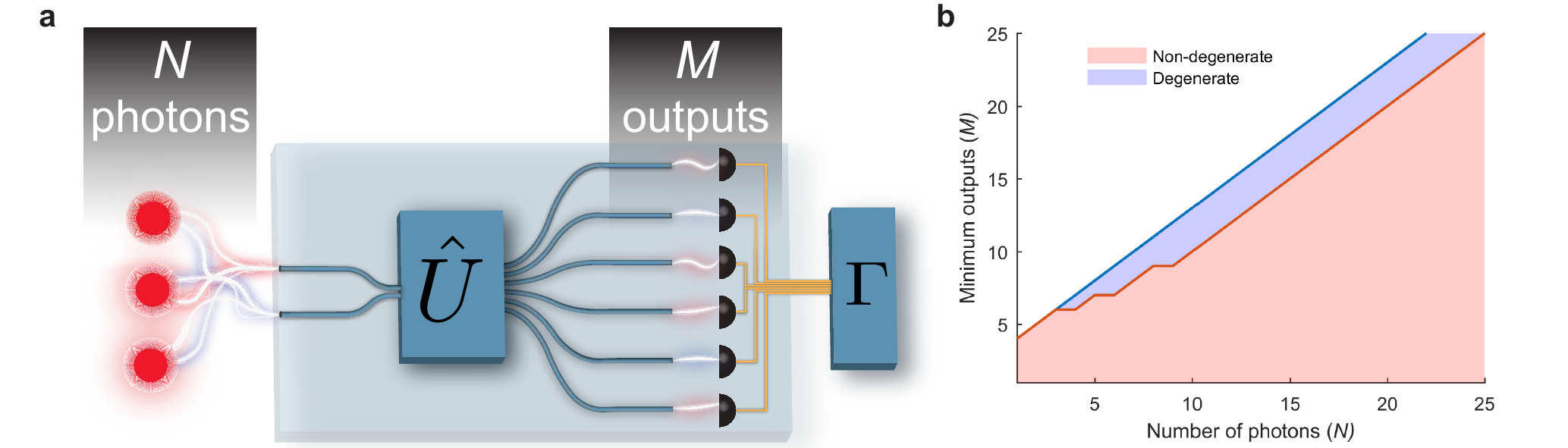}
	\caption{\textbf{Linearly scalable quantum state tomography concept.}
		\textbf{a}, Conceptual diagram of a photonic chip for scalable tomography based on a single unitary transformation. \textbf{b}, Scaling relationship between the number of photons in the quantum state to be measured and the number of output waveguides required in the linear transformation. The scaling for both distinguishable and indistinguishable photons are both considered.
	}
	\label{fig:1}
\end{figure*}

\begin{figure*}
	\centering
	\includegraphics[width=\textwidth]{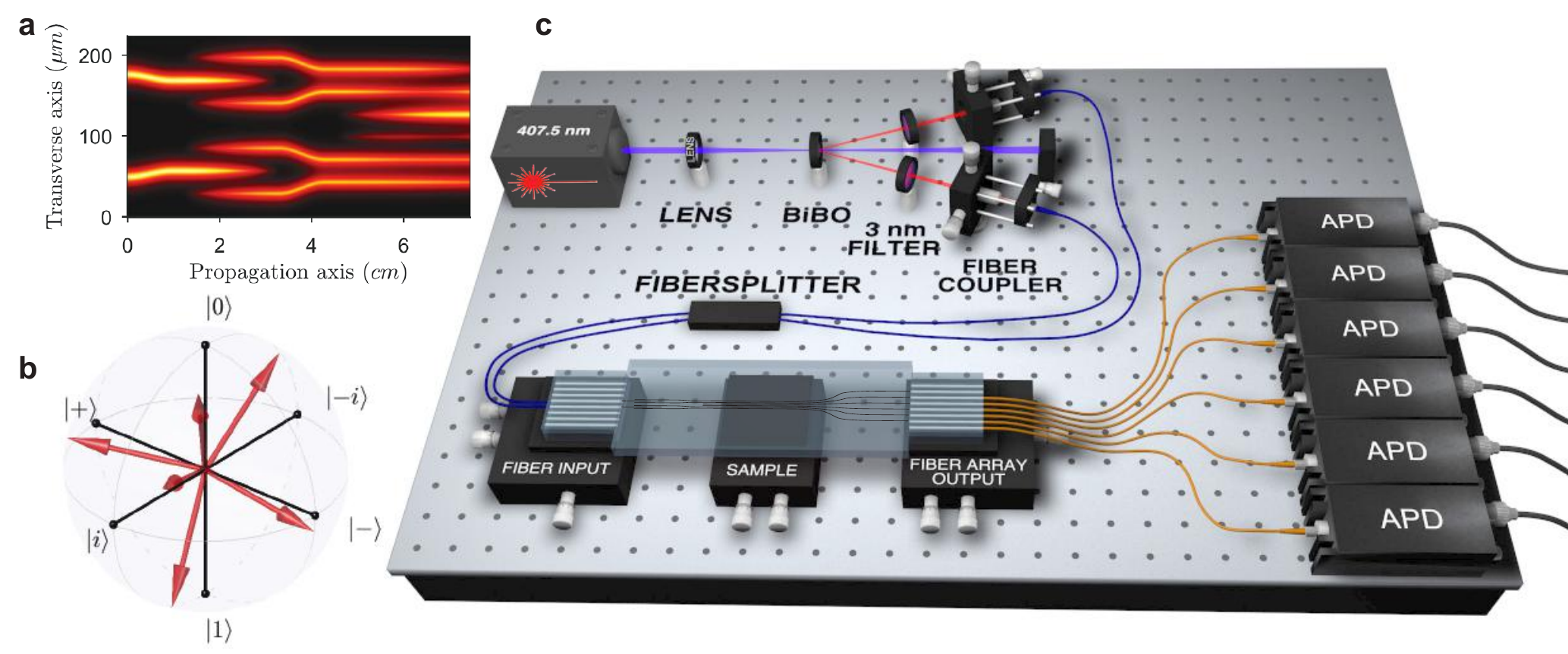}
	\caption{\textbf{Experimental realization with a photonic chip.}
		\textbf{a}, The structure of the silica photonic chip showing simulation of coupling a single photon in state $\ket{0} + i \ket{1}$ into the input ports. \textbf{b}, Experimentally determined mappings from the input Bloch sphere to the six output waveguide field intensities. \textbf{c}, The experimental setup. Photon-pairs at $815$nm are generated via spontaneous parametric down-conversion (SPDC) by pumping a bismuth triborate (BiBO) crystal with a 407.5nm diode laser. The two photons are then coupled into two fibers, and optionally passed through a fiber-splitter to transform the anti-bunched state into a bunched state. The photons are then coupled to the photonic chip and detected with a array of 6 single photon-detecting avalanche photo-diodes (APD's).
	}
	\label{fig:2}
\end{figure*}

Here we present a quantum tomography method with complexity that scales linearly with qubit number. This is achieved by leveraging quantum systems' greatest strength, the simultaneous occupation of exponentially many states, in the measurement process.
%
%\textcolor{red}{this is not a good closing of the paragraph. In the next paragraph we are going to tell that our approach is superior, however this paragraph ends with saying that other approaches are fine as well. We have to make very clear what is the caveat of other approaches and that we provide a solution. Currently, this paragraph sounds way too positive; it needs to tell what is NOT possible until now. In other world: it has to tell that we provide a solution to something no one else found a solution for}
%
Instead of preforming a sequence of different measurements on the state, we design a single static measurement system [Fig.~\ref{fig:1}(a)] that preforms one, many-outcome measurement. Since the state is spread coherently across all the outputs, the number of different measurement outcomes follows a similar scaling to the number of parameters in the density matrix. 
% Instead of preforming a sequence of different measurements on the state, we design a single static measurement system [Fig.~\ref{fig:1}(a)] that can extract all the information required to reconstruct the states density matrix.
%Essentially the sequence of different measurements required in traditional tomography is replaced with a single, many-outcome measurement.
%Intuitively one might expect that the complexity of the apparatus required would scale exponentially with the number of quibts in a similar way to conventional tomography,
Thus the exponential scaling of quantum sates can be balanced by a similar scaling in the amount of information extracted from measurement of the state.
This leads to the striking benefit that the required physical complexity of multi-outcome measurement only scales linearly with the number of qubits in the state being measured, in contrast to the usual exponential scaling.

In the context of photonic quantum states our approach also removes the need to build complex reconfigurable measurement systems \cite{James:2001-52312:PRA}, instead allowing full quantum tomography with just static linear optical circuits, which can easily be implemented on photonic chips.
This avoids the problem faced by conventional approaches, where some measurements provide little useful information about the underlying state. This is because in our approach, the full complement of measurements are performed simultaneously, and thus the most important correlation detections for reconstructing a particular state naturally have the highest count rates.
Our approach is based on interfering all the photons through a special unitary transformation, thus can be optimized to incorporate nonlocal measurement, allowing the error robustness to be increased.
Furthermore, we show that our approach is compatible with computationally scalable reconstruction, avoiding resource intensive direct inversion.

\begin{figure*}
	\centering
	\includegraphics[width=\textwidth]{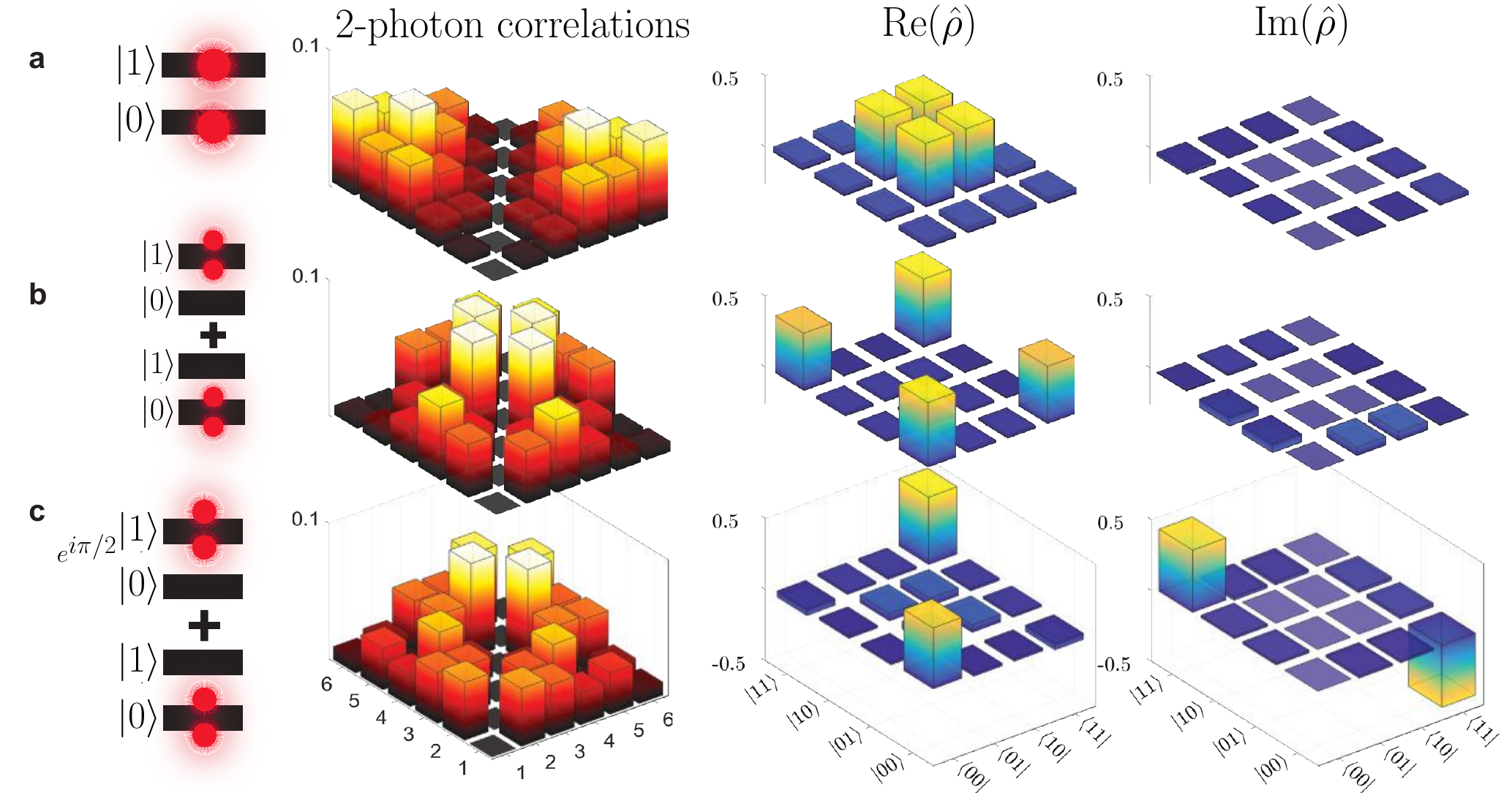}
	\caption{\textbf{Two photon tomography results.}
		\textbf{a}, Results for an anti-bunched state, \textbf{b}, a N00N state, and \textbf{c}, a N00N state with $\pi/2$ phase shift. The second column shows the measured 2-photon correlations, while the third and fourth columns show the real and imaginary parts of the recovered density matrices.
	}
	\label{fig:3}
\end{figure*}

 We explain our method for the case of $N$-photons in an arbitrary pure or mixed state featuring spatial quantum entanglement between two input ports, although a larger number of ports can be considered as well \cite{Titchener:2016-4079:OL}.
 The $N$-photon quantum state is described by a density matrix, $\hat{\rho}_{in}$, which has $2^{2N}$ real parameters. We apply a single (fixed) linear transformation, $\hat{U}$, to map the quantum state from two inputs to a larger number of output waveguides, $M$, as illustrated in Fig.~\ref{fig:1}(a).
 Then, we obtain information about the quantum state by measuring coincidences, $\Gamma$, in the arrival time of photons to different combinations of the single-photon detectors.
 There are $M!/(M-N)!$ different $N$-photon correlations which can be measured, assuming the detectors cannot resolve photon numbers, but can distinguish photons (i.e. by arrival time for time-bin encoded states).
 The reconstruction of the density matrix is possible if the number of different correlations is larger than the number of unknown density matrix elements.
% For the most general case, where the $N$ photons have varying amounts of distinguishably to one another, the condition is satisfied if $M>N$ (when the number of photons is greater than $10$). therefore the number of output ports required scales linearly with the number of photons as shown in Fig.~\ref{fig:1}(b).
We establish that the minimum number of output ports required scales linearly with the number of photons and approaches
%This condition is satisfied if
$M=N+3$ for large photon numbers, both in cases of distinguishable (red shaded area) and indistinguishable (blue line) photons as shown
%as shown by the red shaded area
in Fig.~\ref{fig:1}(b); mathematical details are provided in the Sec. 2 of the supplementary information.

%In this work we consider the case where the measurement apparatus does not distinguish which photon is detected, and furthermore we assume the quantum state to be degenerate, meaning that the photons are all completely indistinguishable. In this case the scaling of our method remains linear as shown by the blue line in Fig.~\ref{fig:1}(b). Finally in the supplementary material, we show that even without photon distinguishing detectors our method can in-fact recover parts of the density matrix related to the photon distinguishably, so is also applicable to non-degenerate photon states.

%In the case where the the measurement apparatus does not distinguish which photon is detected the number of distinct types $N$ photon correlation that can be measured is reduced to $M!/N!/(M-N)!$. Assuming that the $N$ photons are degenerate the number of degrees of freedom in the density matrix reduces comparatively to $(N+1)^2$ to preserving the linear scaling of the number of waveguides required, as shown by the blue line in \ref{fig:1}(b). This is the case we consider experimentally in this work. However in fact even in the case of non-degenerate photons our method can still recover most parts of the density matrix without need for photon distinguishing measurements. In particular we can recover information about the distinguishably of the photons as explained in the supplementary material.

We experimentally demonstrate our approach by performing tomography of spatially entangled mixed or pure states of two indistinguishable photons,
%In this work we consider the case where the measurement apparatus does not distinguish which photon is detected, and furthermore we assume the quantum state to be degenerate, meaning that the photons are all completely indistinguishable.
using a specially designed on-chip laser-written waveguide circuit \cite{Meany:2015-363:LPR}. The action of the circuit on the input single photon state $\ket{\psi} = \ket{0}+ i\ket{1}$ is shown in  Fig.~\ref{fig:2}(a). The circuit allows full reconstruction of the input density matrix just by measuring the output two-photon coincidences with non-photon-number resolving single-photon detectors.

The circuit was optimized to make the tomographic reconstruction highly robust to measurement errors. Each output waveguide carries information from a different vector on the input Bloch sphere. The Bloch vectors were determined experimentally using a classical characterization method \cite{Heilmann:2015-96:RAR}, and are shown graphically in Fig.~\ref{fig:2}(b). The equal spacing of vectors around the Bloch sphere gives the device maximum robustness to errors in the tomography procedure \cite{Foreman:2015-263901:PRL}, by essentially realizing non-local measurements. This can be confirmed by calculating the condition number of the transfer function of the chip, lower values of which correspond to higher robustness of the state reconstruction to measurement errors. The experimentally realized device has a condition number of $\simeq5$, which is better than the condition number of $\simeq9$ for typical to tomography \cite{Miranowicz:2014-62123:PRA}.

To test the performance of the device we prepared a range of different two-photon entangled quantum states, coupled them into the chip, and measured the output correlations as schematically shown in Fig.~\ref{fig:2}(c). We first analyze an anti-bunched state, which in an ideal form is described by the pure wavefunction $\ket{\psi} = \left( \ket{0 1} + \ket{1 0} \right)/\sqrt{2}$ and the corresponding density matrix $\hat{\rho} = \ket{\psi}\bra{\psi}$.
We present in Fig.~\ref{fig:3}(a) the experimentally measured probabilities of detecting the photon-pair in a given pair of output waveguides, and the reconstructed
%for an input anti-bunched state. Such an ideal pure state is described by the wavefunction $\ket{\psi} = \left( \ket{0 1} + \ket{1 0} \right)/\sqrt{2}$ and the corresponding density matrix $\hat{\rho} = \ket{\psi}\bra{\psi}$.
%The
real and imaginary parts of the density matrix, as indicated by labels.
%were reconstructed from the correlation measurement and are shown in Fig.~\ref{fig:3}(a) in columns 3 and 4, respectively.
We confirm that this is indeed an anti-bunched state, with the fidelity of $95.0\%$. Furthermore, this measurement permits us to get information about the spectral overlap of the pair of photons, since the observed correlations exhibit a generalized form of Hong-Ou-Mandel interference \cite{Hong:1987-2044:PRL,Spring:2017-90:OPT, Mahrlein:2015-15833:OE}, see Sec. 1 of the supplementary material for details.

We also prepared N00N states, with wavefunctions given by $\ket{\psi} = \left( \ket{0 0} + e^{i\phi} \ket{1 1} \right)/\sqrt{2}$, where $\phi$ is a phase shift. The phase shift is determined by the photon propagation before the chip, and because the shift is double the value that would accumulate classically~\cite{Ou:1990-2957:PRA} it is highly sensitive to the environment. Experimentally, we explored this quantum-enhanced sensitivity by propagating a two-photon N00N state through one meter long optical fibers before the chip. The accumulated phase was very sensitive to fiber stress, varying by up to $2\pi$ on the scale of a few minutes.
With our approach we observed experimentally the temporal variation of the phase in the density matrix using an integration time of $20$ seconds with a photon pair detection rate of $30$Hz.
We show typical two-photon correlations at different times in Figs.~\ref{fig:2}(b) and (c) and the corresponding reconstructed N00N states, with phases determined to be $\phi=0$ in Fig.~\ref{fig:3}(b) and $\phi = \pi/2$ in Fig.~\ref{fig:3}(c). The fidelity of both states exceeds $94\%$. Thus,
%with our approach to quantum tomography
we can observe with high precision the density matrix of a quantum state that is varying over time (a video of the time evolution of the density matrix is included in the supplementary material).

\begin{figure}
	\centering
	\includegraphics[width=\columnwidth]{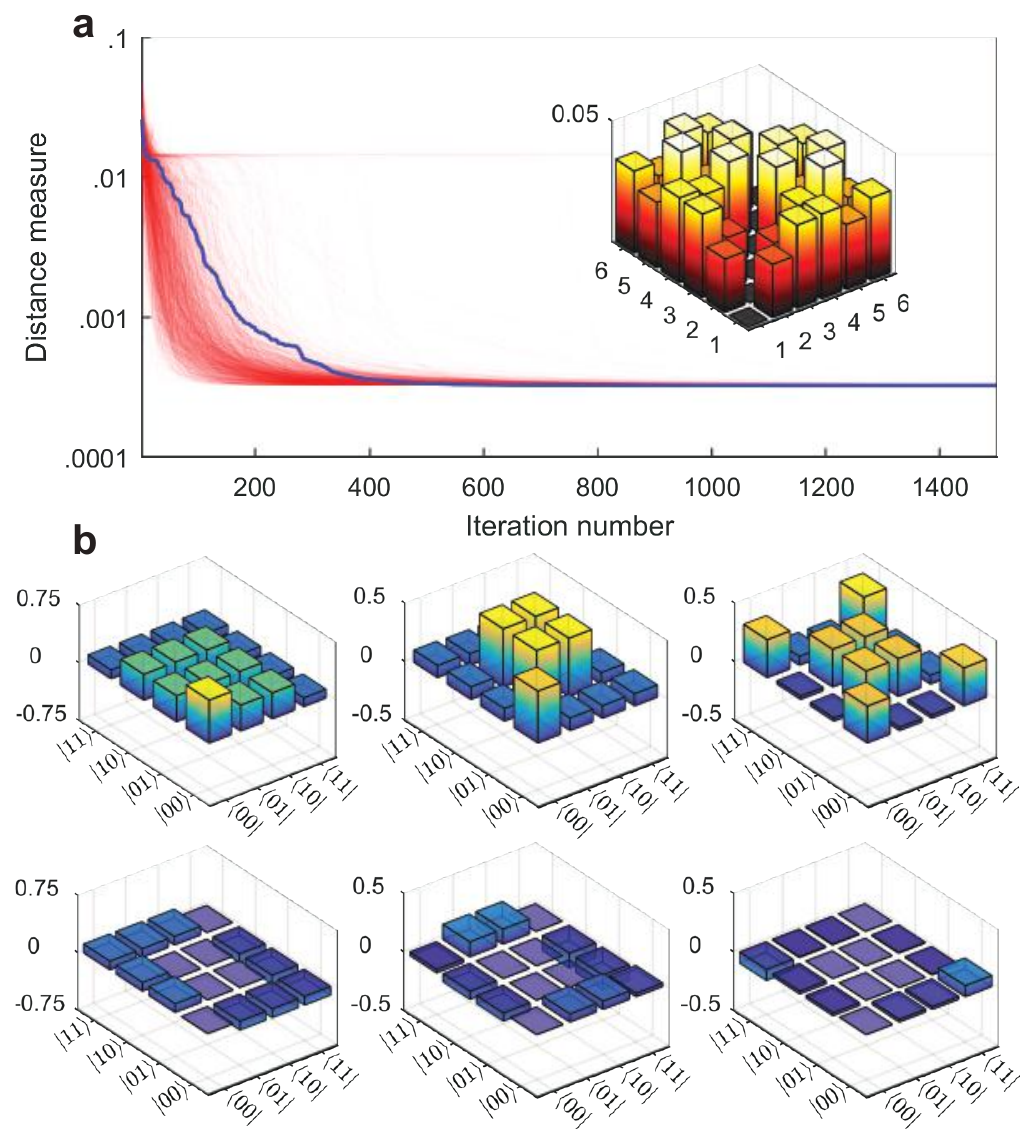}
	\caption{\textbf{Scalable reconstruction algorithm for mixed states.}
		\textbf{a}, \textit{inset} Measured two-photon correlations for the mixed state $\hat{\rho}_\text{mix}  = \left(\hat{\rho}_\text{anti-bunched}+\hat{\rho}_\text{N00N} \right)/2$. \textbf{a}, Performance of the self guided tomography algorithm searching for the input density matrix that best matches the measured correlations. Red curves show 1000 different realizations of the algorithm. \textbf{b}, The real (top row) and imaginary (bottom row) parts of the density matrix at iteration numbers 1, 30 and 300 during the realization highlighted in blue in Fig.~\ref{fig:4}(a). 	}
	\label{fig:4}
\end{figure}

\begin{figure}
	\centering
	\includegraphics[width=\columnwidth]{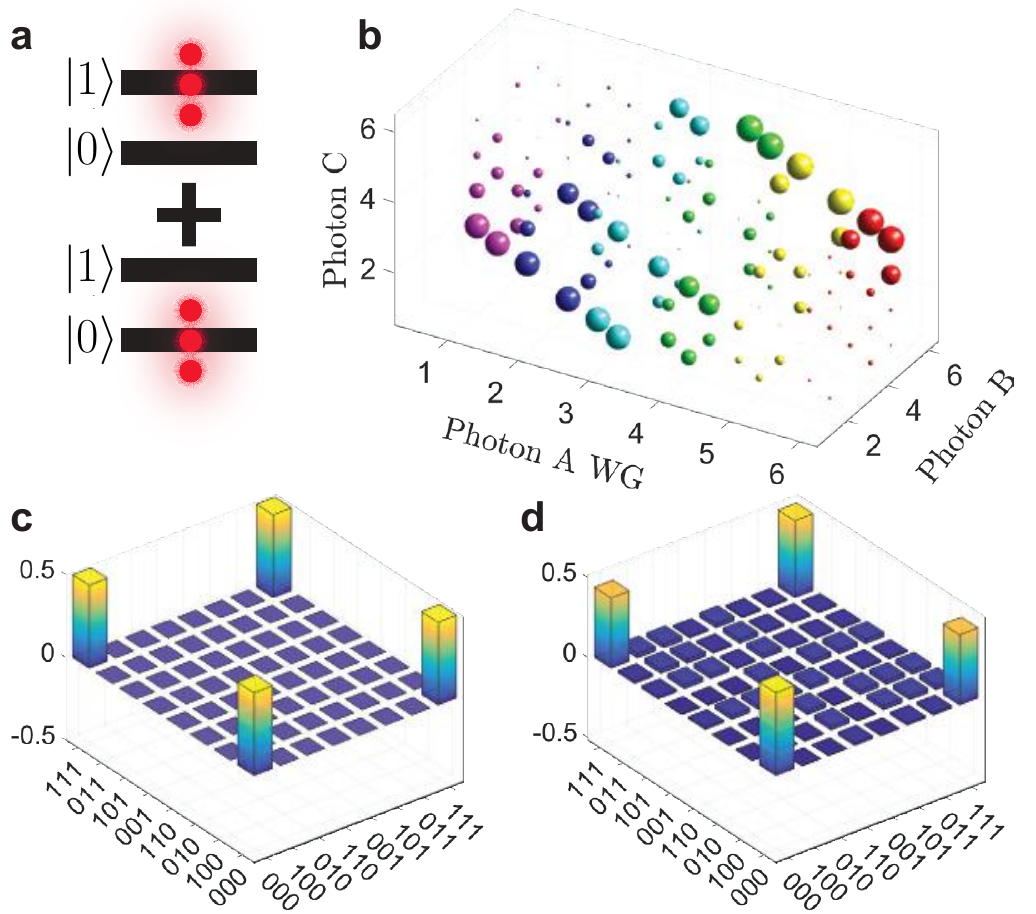}
	\caption{\textbf{Tomography of a three-photon state.}
		\textbf{a}, Diagram of the three-photon GHZ state. \textbf{b}, Simulated correlations of the GHZ state after propagation through the measured transfer function of the device in Fig.~\ref{fig:2}(a). Gaussian error with standard deviation of 5\% the maximum correlation element's value is added to each element. \textbf{c}, The real part of the GHZ state's density matrix. \textbf{d}, Real part of the  density matrix that was recovered using the simulated correlations from (b).
	}
	\label{fig:5}
\end{figure}

The reconstruction of the density matrices from the measured correlations was carried out using a computationally scalable algorithm.
This is important, since although the number of detectors and waveguides in the circuit scales linearly with the number of qubits to be measured, the number of photon detections required for reconstruction still necessarily scales exponentially, and processing of measurement data can be extremely resource demanding.
We employ an optimization technique known as the simultaneous perturbation
stochastic approximation \cite{Spall:1992-332:RAR}, similar to the algorithm formulated for self-guided quantum state tomography \cite{Ferrie:2014-190404:PRL}. It minimizes a distance measure between the true state and the algorithm's current guess. We use the least squares distance between the measured correlations and the correlations that would be produced by the current guess of the density matrix, $|\Gamma_{\text{meas.}} - \Gamma_{\text{guess}}|^2$. The reconstruction fidelity of the algorithm is $99.67\%$ after 1500 iterations (see Methods).

We demonstrate recovery of the mixed state $\hat{\rho}_\text{mix}  = \left(\hat{\rho}_\text{anti-bunched}+\hat{\rho}_\text{N00N} \right)/2$
using the algorithm.
The inset of Fig.~\ref{fig:4}(a) shows the experimentally measured correlations corresponding to $\hat{\rho}_{\text{mix}}$, obtained by numerically combining separately recorded raw coincidence data from an anti-bunched state and a N00N state. The main plot in Fig.~\ref{fig:4}(a) shows 1000 realizations of the algorithm as red lines, and density matrices at iteration numbers 1, 30 and 300 are shown in Fig.~\ref{fig:4}(b) for the realization highlighted in blue in Fig.~\ref{fig:4}(a).
%We observe a robust convergence to the expected state. The  showing
Importantly, our method offers the same computational advantage as self-guided tomography, but without a need for complex reconfigurable measurements which so far restricted this approach to only pure states~\cite{Chapman:2016-40402:PRL}.

%The progress of the algorithm minimizing the distance measure is shown in Fig.~\ref{fig:4}(a) while the inset shows the measured correlations it is optimizing to fit.
%The figure shows 1000 different iterations of the algorithm as red lines, demonstrating the convergence to a particular density matrix. The density matrix at iteration numbers 1, 30 and 300 is shown in Fig.~\ref{fig:4}(b) for the realization highlighted in blue in Fig.~\ref{fig:4}(a).
%The algorithm converges to the expected mixed state $\hat{\rho}  = \left(\hat{\rho}_{anti-bunched}+\hat{\rho}_{N00N} \right)/2$ showing our algorithm can recover quantum mixed states. It should be noted, that although the number of detectors and waveguides in the circuit scales linearly with the number of qubits in the state to be measured, the number of photons required to be detected for reconstruction still necessarily scales exponentially.

The chip presented in Fig.~\ref{fig:2}(a) is also capable of tomography of degenerate three-photon states. We demonstrate this using the experimentally determined transfer function, $\hat{U}$, and simulate the propagation of a three-photon Greenberger-Horne-Zeilinger (GHZ) state [Fig.~\ref{fig:5}(a)] through the chip. The simulated output three-photon correlations are shown in Fig.~\ref{fig:5}(b), where Gaussian noise with standard deviation $5\%$ of the peak correlation value has been added to each element. Reconstruction of the input density matrix gives highly accurate results despite this noise. The real part of the recovered density matrix is shown in Fig.~\ref{fig:5}(d), which closely matches the three-photon GHZ state in Fig.~\ref{fig:5}(c). This provides a significantly simpler and more stable platform for three-photon tomography compared to previous realizations \cite{Hamel:2014-801:NPHOT}.

In our work, we have demonstrated that a fixed linear optical transformation can be devised to allow complete quantum tomography of $N$-photon states. Importantly, the complexity of the transformation only scales linearly with the number of photons in the state, in contrast to the exponential scaling in the number of measurements required in usual tomography \cite{White:1999-3103:PRL}. Due to its simplicity and lack of tunable elements, this approach to quantum measurement is uniquely suited to integration with on-chip single photon detectors \cite{Najafi:2015-5873:NCOM} for a fully on-chip tomography scheme. This provides a promising way to facilitate the characterization and development of increasingly complex quantum communication and computation systems.    

%In our work, we have demonstrated that fixed linear optical transformations can be devised to allow complete quantum tomography of $N$-photon states. This approach permits non-local measurement, and the exact form of the transformation can be optimized to give more error robust measurements than conventional tomography. Importantly, the complexity of the transformation only scales linearly with the number of photons in the state to be measured, in contrast to the exponential scaling in the number of measurements required in usual tomography \cite{White:1999-3103:PRL}. Due to its simplicity and scalability, this approach to quantum measurement is uniquely suited to integration with on-chip single photon detectors \cite{Najafi:2015-5873:NCOM} for a fully on-chip tomography scheme. This linearly scalable tomography with integrated multi-outcome measurements provides a promising way for quantum measurement to keep pace with the rapidly increasing complexity of quantum devices.    

%In particular the scheme could be generalized for tomography of states with unknown photon number, for example, by simultaneously measuring $2$-, $3$- and $4$-photon correlations. \textcolor{red}{There is some inspirational sentence in the end missing: where does it lead to, what cool stuff might this allow in the future?}

%cite{Weingartner1970}
%\bibliography{Tomography Paper}

%%%%%%%%%%%%%%%%%%%%%%%%%%%%%%%%%%%%%%%%%%%%%%%%%%%%%%%%%%%%%%%%%%%%%%%%%%%%
%%%%%%%%%%%%%%%%%%%%%%%%%%%%%%%%%%%%%%%%%%%%%%%%%%%%%%%%%%%%%%%%%%%%%%%%%%%%

\section{Methods}

\textbf{Direct waveguide writing.} We write our waveguides into transparent fused silica wafers (Corning 7980 ArF Grade), using ultrashort laser pulses ($\tau<150$\,fs, $\lambda=800$\,nm) that are focused 250\,$\mu$m below the sample surface using a 20$\times$ microscope objective (NA\,$ \approx$\,0.35). The actual writing speed, achieved with a high-precision positioning system (Aerotech) is 100\,mm/min at a pulse energy of 200\,nJ and a repetition rate of 100\,kHz (Coherent Mira/Reg A). Such waveguides exhibit low propagation losses (\,$<$\,$0.3$\,dB/cm) and almost vanishing birefringence ($\Delta n_{H,V} \approx 10^{-6}$). With a supported mode field diameter of $12$\,$\mu$m\,$\times$\,$15$\,$\mu$m coupling losses of $3$\,dB are obtained with standard single mode fibers.

\textbf{Photon-pair generation and measurement.} We generate photon-pairs at $\lambda=815$\,nm using a standard type-I spontaneous parametric down-conversion source with a visibility of 93\%. A BiBO crystal is pumped by a $100$\,mW, $407.5$\,nm laser diode producing horizontally polarized photon-pairs, which are collected by polarization maintaining fibers. Commercial V-groove fiber arrays were used to couple the photons into the chip as well as collecting them at the output facet from the individual waveguides. We used high-NA multi-mode fibers in order to feed the photons to the respective avalanche photo-diodes, ensuring low coupling losses at the output side of the chip. From the data of the photo-diodes, the photon probability distribution at the output, as well as the inter-channel correlations, were computed using a correlation device (Becker-Hickl) and standard computer programs (LabView for the data acquisition and MatLab for the data processing). The photonic chip was fabricated such that the waveguide spacing at the input and output facets matched the standard fiber array spacing of $127\,\mu$m. The typical two-photon coincidence rates after propagating through the chip were approximately $30$Hz, and an integration time of around $20$ seconds was used for coincidence measurements.

\textbf{Algorithmic reconstruction.}
The algorithm for reconstructing the density matrix from a measured set of correlations uses the simultaneous perturbation
stochastic approximation (SPSA) \cite{Spall:1992-332:RAR}. 
We follow an optimization process very similar to \cite{Ferrie:2014-190404:PRL}, except we adapt it to work with mixed quantum states.
%It works starting from a randomly chosen `guess' of the system's density matrix then proposing a random direction to search in, $\hat{\sigma}$, throughout the space of physical density matrices.
This is achieved by defining the density matrix as, $\hat{\rho}(t)=\hat{T}^{\dagger}(t)\hat{T}(t)/\text{Tr}\{\hat{T}^{\dagger}(t)\hat{T}(t)\}$, as in Eq.~(4.5) of Ref.~\cite{James:2001-52312:PRA}. Then the optimization algorithm proceeds optimizing $t$, rather than $\psi$ as in previous approaches for pure quantum states. 
%The algorithm starts from a random physical density matrix, $\hat{\rho}(t_1)$ and proposes two new density matrices given by $\hat{\rho}(t_1 \pm \sigma)$, where $\sigma$ is a randomly generated search direction. The next iteration of the density matrix is updated according to $\hat{\rho}(t_2) = \hat{\rho}(t_1+g\sigma)$, where $g$ is the difference between the distance measures evaluated for the two proposed density matrices. 
We determine the reconstruction accuracy using the {\em experimentally measured} device transfer function. The recovery of 500 random density matrices is simulated, each for 1500 iterations of the algorithm. The reconstruction fidelity, the average fidelity between the true density matrix and the algorithm's final guess, is found to be $99.67\%$.

%to be closer to the proposed density matrix, depending on the least squares distance between the measured correlations and the correlations that would be produced by the current guess of the density matrix, $|\Gamma_{\text{meas.}} - \Gamma_{\text{guess}}|^2$.

\section{Acknowledgments}
%(Other grants?
%)
%Nano-Phi project for travel funding
We acknowledge support by the Australian Research Council (ARC) (DP130100135, DP160100619); Erasmus Mundus
(NANOPHI 2013 5659/002-001); Alexander von Humboldt-Stiftung; Australia-Germany Joint Research Co-operation
Scheme of Universities Australia; German Academic Exchange Service.

\section{Author contributions}
J.G.T, A.S.S, and A.A.S developed the theoretical concept. J.G.T, M.G, R.H, and A.S designed and fabricated the photonic chip. J.G.T, M.G, and R.H preformed the experiments. All authors co-wrote the manuscript.

%%%%%%%%%%%%%%%%%%%%%%%%%%%%%%%%%%%%%%%%%%%%%%%%%%%%%%%%%%%
\section{Supplementary material}

     \subsection{Classical characterization of the device transfer function}

     \begin{figure}
     	\centering
     	\includegraphics[width = \textwidth]{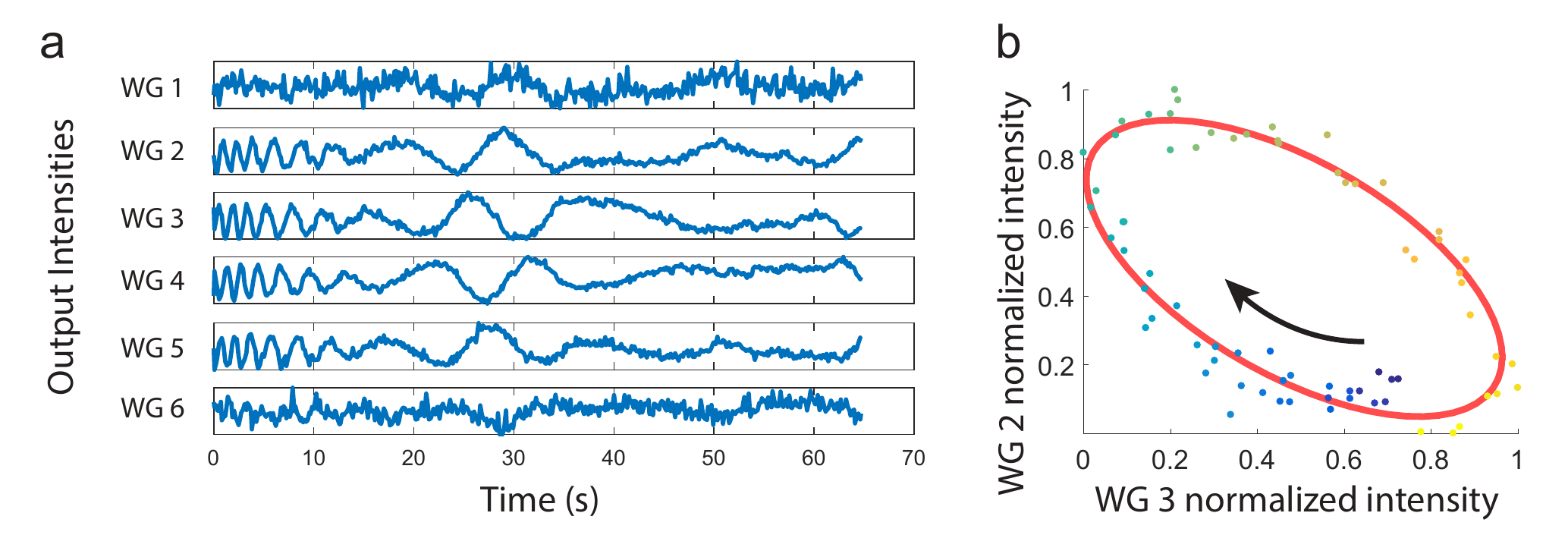}
     	\caption{\textbf{ Measurement of transfer function elements phases.} (a) Output intensities of each waveguide when a $810\,nm$ laser is coupled into each of the two input ports simultaneously using a beamsplitter. The intensity of each waveguide is normalized so that the maximum intensity value is 1 for each waveguide.  (b) Colored data points show the intensity in waveguide 2 against the intensity in waveguide 3. The time each data point is recoreded is shown by its color, with the earliest data points being blue.
     	}
     	\label{fig:4}
     \end{figure}

     Using the photonic chip for tomography requires that the action of the chip on an input photon can be accurately predicted. The propagation of a photon wavefunction through the chip can be described by $\mathbf{\hat{U}} \ket{\psi_{in}} = \ket{\psi_{out}}$, here $\mathbf{\hat{U}}$ is the transfer function of the chip, describing the mapping from the input waveguides to the outputs. Thus in order to use the chip for tomography this transfer function needs to be determined. This could be achieved using quantum process tomography, however we used a more straightforward method \cite{Heilmann:2015-96:RAR} requiring just classical light.

     We use a classical laser beam coupled into the chip in order to find its transfer function. The absolute values of each element in the transfer function, $|U_{i,j}|^2$ are found simply by coupling the laser into one of the inputs and measuring the intensity at each of the 6 outputs. For example coupling the laser into input waveguide $1$ and measuring the intensity at output waveguide 6 will give the transfer element $|U_{1,6}|^2$. Doing this for both the two input ports of the chip gives the absolute values of each of the $12$ elements in the transfer function.

     In a similar way the phases of the transfer function can be easily determined, up to a constant overall phase factor between the two columns of the transfer function.
     %We write the complex transfer function as $|U_{j,k}|e^{i\phi_{j,k}}$ and
     To do this we pass the laser through a beamsplitter, splitting it into two beams, then couple each beam into one of the two input waveguides. The two beams interfere within the chip. The output field in the $j^{\text{th}}$ waveguide is be given by,
          \begin{equation}
          U_{1,j}\,E_{1}^{(in)} + U_{2,j}\,E_{2}^{(in)} = E_{j}^{(out)}
          \label{map}
          \end{equation}
Here $E_{1}^{(in)}$ is the complex field amplitude at input waveguide number $1$ and $E_{2}^{(in)}$ is the complex field amplitude at input $2$.
Since the absolute values of the transfer matrix elements are already known, it makes sense to separate the transfer matrix into its absolute values and phase components.  Thus we rewrite Eq.~(\ref{map}) as,
     \begin{equation}
     \left|	|U_{1,j}||\,E_{1}^{(in)}| + |U_{2,j}|\,|E_{2}^{(in)}|e^{i(\phi_j+\Delta d/\lambda)} \right|^2 = |E_{j}^{(out)}|^2
     	\label{interference}
     \end{equation}
Here $\phi_j$ is the phase difference between $U_{1,j}$ and $U_{2,j}$, which we need to determine. The global phase $\Delta d/\lambda$ denotes a variable phase shift we will intentionally introduce by increasing the path length of the beam coupled into input 2.
Varying the phase of one input relative to the other is ultimately what will allow the unknown transfer matrix phases, $\phi_j$, to be determined. The changing phase produces oscillations in the intensity of each output waveguide. These oscillations measured in our chip are shown in Fig.~\ref{fig:4}(a). The output intensity,  $|E_{j}^{(out)}|^2$, will oscillate with changing $\Delta d$ according to

          \begin{equation}
           |E_{j}^{(out)}|^2  = \frac{a}{2}[1 + \cos({\phi_{j}+\Delta d/\lambda)}] + b
          \end{equation}

    Here $a$ and $b$ are constants related to absolute values of the transfer matrix and input beams. Since $\Delta d/\lambda$ at a given time is the same for all output ports, the phase shifts between transfer matrix elements, $\phi_{j}$, can to be determined, up to an unknown phase shift, $\theta$, between the two columns of the transfer function. This is achieved by comparing the phase shifts between the oscillations of the different outputs in Fig.~\ref{fig:4}(a).
    To determine the phase shifts efficiently we plot the output intensity of waveguide $j$ against the intensity of waveguide $k$, which yields elliptical Lissajous curves. The shape and orientation of the ellipses' determine the phase difference between the oscillations of the intensity in waveguides $j$ and $k$. For example the Lissajous curve for waveguides $2$ and $3$ is plotted in Fig.~\ref{fig:4}(b). An ellipse is fitted to the data points and the parameters of the ellipse determine the phase difference between the oscillations in outputs $2$ and $3$, thus give the quantity $\phi_2-\phi_3$. This is repeated for all combinations of output ports to find all the phases of the transfer function, up to an unknown constant phase shift $e^{i\theta}$ between the two columns of the transfer function, as in \cite{Heilmann:2015-96:RAR}.

     After extracting the phases between all the elements in one column of the transfer function we can now write the transfer function up to the constant overall phase factor $e^{i\theta}$,
     \begin{equation}
     	|U_{1,j}|\,E_{1}^{(in)} + |U_{2,j}|e^{i(\phi_{j}+\theta)}\,E_{2}^{(in)}  = E_{j}^{(out)}
     \end{equation}
     This constant phase factor, $\theta$, is unknown because the absolute value of the phase shift between the two inputs $E_1^{(in)}$ and $E_2^{(in)}$ is never directly measured, all that is known is that the shift is changing with the displacement $\Delta d$.
This unknown phase factor could have been determined using an interferometric setup, but we chose not to do this since we can already observe relative phase differences between different density matrices without needing to know the absolute phase.

%%%%%%%%%%%%%%%%%%

%%%%%%%%%%%%%%%%%%

%%%%%%%%%%%%%%%%%%

\subsection{The spatial density matrix of a broadband two-photon wavefunction}
In the main work, `Scalable on-chip quantum state tomography', we characterized the spatial density matrix of a number of different two-photon quantum states. Of course these states also have spectral and temporal properties, and these properties have an impact on the spatial density matrices we recover. In this section we show a link between the spectral/temporal properties of the photon-pair and the resulting spatial density matrices.
We do this starting from a general frequency dependent two-photon wavefunction of the type that would be produced by our SPDC source, then we integrate over the photon frequencies in order to derive the reduced density matrix of the spatial state of the two-photons.
Finally we link this analytic result to real measurements of two-photon states with different temporal delays, demonstrating that our approach to tomography can recover information about spectral/temporal distinguishably of photons.

First we write the frequency dependent two-photon wavefunction as

\begin{equation}
\ket{\Psi} =
\sum_{n,m} \int d\omega_1 d\omega_2 \,
\, \Psi_{n,m}(\omega_1,\omega_2) \,
\hat{a}_n^{\dagger}(\omega_1)\hat{a}_m^{\dagger}(\omega_2)\ket{0}
\end{equation}

Here  $\psi_{n,m}(\omega_1,\omega_2)$ is the joint spatial and spectral distribution of photons $a$ and $b$. The spatial mode of photon $a$ is denoted $n$ and the mode of photon $b$ is denoted $m$.

For comparison with experiment we now specify the form of the wavefunction to be an anti-bunched state, of the type that would be produced by our SPDC source. Specifically we set
\begin{equation}
\Psi_{n,m}(\omega_1,\omega_2) = \phi^{(a)}(\omega_1)\phi^{(b)}(\omega_2)\delta_{n,1}\delta_{m,2}
\end{equation}
Giving photons $a$ and $b$ spectra $\phi^{(a)}$ and $\phi^{(b)}$, and putting them in different spatial modes to make an anti-bunched state.

Elements in the reduced spatial density matrix $\hat{\rho}_{n,m,n',m'}$ can be calculated by projecting the two-photon wavefunction into the operator basis for the spatial density matrix, effectively integrating over frequency to leave a purely spatial description of the density matrix. So we will take inner products with the following operators,

\begin{equation}
	\ket{\boldsymbol{D}_{n,m}} \bra{\boldsymbol{D}_{n',m'}}  = \\
	\int d\omega_3 d\omega_4 \,
	D^\ast(\omega_3) \, F^\ast(\omega_4)\hat{a}_n^{\dagger}(\omega_3)\hat{a}_m^{\dagger}(\omega_4)\ket{0}
	\bra{0}
	D(\omega_3) \, F(\omega_4)\hat{a}_{n'}(\omega_3)\hat{a}_{m'}(\omega_4),
\end{equation}

Where $D$ and $F$ represent some spectral or temporal windows used to distinguish the two photons. In the case where the measurement apparatus is not designed to distinguish different photons we set $D^\ast(\omega_3)D(\omega_3)= F^\ast(\omega_3)F(\omega_3) = 1/\sqrt{2}$, to preserve normalization since each photon will be detected by both detection windows. Otherwise $D$ and $F$ can be set to be step functions in frequency or time, depending of whether photons are distinguished by their spectrum or arrival times.

So the elements of the anti-bunched state's density matrix are given by the expectation values,
\begin{multline}
	\hat{\rho}_{n,m,n',m'} = \bra{\Psi}\ket{\boldsymbol{D}_{n,m}} \bra{\boldsymbol{D}_{n',m'}} \ket{\Psi}=
	\\
	%\hat{\rho}_{n,m,n',m'} =
	\frac{1}{2}
	\bra{0}
	\int d\omega_1 d\omega_2 \,
	\phi^{(a)\ast}(\omega_1)\phi^{(b)\ast}(\omega_2)
	\hat{a}_1(\omega_1)\hat{a}_2(\omega_2)
	\times\\
	\int d\omega_3 d\omega_4 \,
	D^\ast(\omega_3) \, F^\ast(\omega_4)\hat{a}_n^{\dagger}(\omega_3)\hat{a}_m^{\dagger}(\omega_4)\ket{0}
	\bra{0}
	D(\omega_3) \, F(\omega_4)\hat{a}_{n'}(\omega_3)\hat{a}_{m'}(\omega_4)
	\times\\
	\int d\omega_5 d\omega_6 \,
	\phi^{(a)}(\omega_5)\phi^{(b)}(\omega_6)
	\hat{a}_{1}^{\dagger}(\omega_5)\hat{a}_{2}^{\dagger}(\omega_6)\ket{0}.
\end{multline}

This can be evaluated to give,

\begin{equation}
\hat{\rho}_{1,2,1,2} =
\int d\omega_3 d\omega_4 \,
\phi^{(a)\ast}(\omega_3)\phi^{(b)\ast}(\omega_4)
\phi^{(a)}(\omega_3)\phi^{(b)}(\omega_4) = 1/2,
\end{equation}
\begin{equation}
\hat{\rho}_{2,1,2,1} =
\int d\omega_3 d\omega_4 \,
\phi^{(a)\ast}(\omega_4)\phi^{(b)\ast}(\omega_3)
\phi^{(a)}(\omega_4)\phi^{(b)}(\omega_3) = 1/2,
\end{equation}
\begin{multline}
\hat{\rho}_{1,2,2,1} = \hat{\rho}_{2,1,1,2}^{\ast} =
\int d\omega_3 d\omega_4 \,
\phi^{(a)\ast}(\omega_3)\phi^{(b)\ast}(\omega_4)
\phi^{(a)}(\omega_4)\phi^{(b)}(\omega_3)/2 = \\
\int d\omega_3 \,
\phi^{(a)\ast}(\omega_3) \phi^{(b)}(\omega_3)
\int d\omega_4
\phi^{(b)\ast}(\omega_4) \phi^{(a)}(\omega_4)/2.
\end{multline}

All other density matrix elements are zero. 
Thus general form of the anti-bunched state density matrix will be,

\begin{equation}
\hat{\rho} =
\frac{1}{2}
\begin{bmatrix}
0     & 0 & 0 & 0 \\
0 & 1   & I_{a,b}I_{b,a} & 0 \\
0 & I_{a,b}I_{b,a}   & 1 & 0 \\
0     & 0 & 0 & 0
\end{bmatrix},
\label{full_D}
\end{equation}
where $ I_{a,b} = \int d\omega\,
\phi^{(a)\ast}(\omega) \phi^{(b)}(\omega)$ is the spectral overlap between the two photons.
Clearly this spectral overlap
is an important factor in determining the exact form of the two photon density matrix. It can be equal to zero in cases where the photons have non-overlapping spectrum or if there is a temporal delay between the photons. In the case where the photons are indistinguishable the overlap will be unity, resulting in a different spatial density matrix compared to the non overlapping (distinguishable case) case.

The reduced density matrix of the anti-bunched state [Eq.~\ref{full_D}] can be recovered using our static photonic chip based approach to tomography. This can be achieved for quantum states of both distinguishable and indistinguishable photons, despite the fact that we do not use single-photon detectors to directly determine which photon was detected. Ultimately this is possible because of generalized Hong-ou-Mandel (HOM)  \cite{Hong:1987-2044:PRL} interference between the photons in the case where their spectra overlap.

        \begin{figure}
        	\centering
        	\includegraphics[width = 0.7\textwidth]{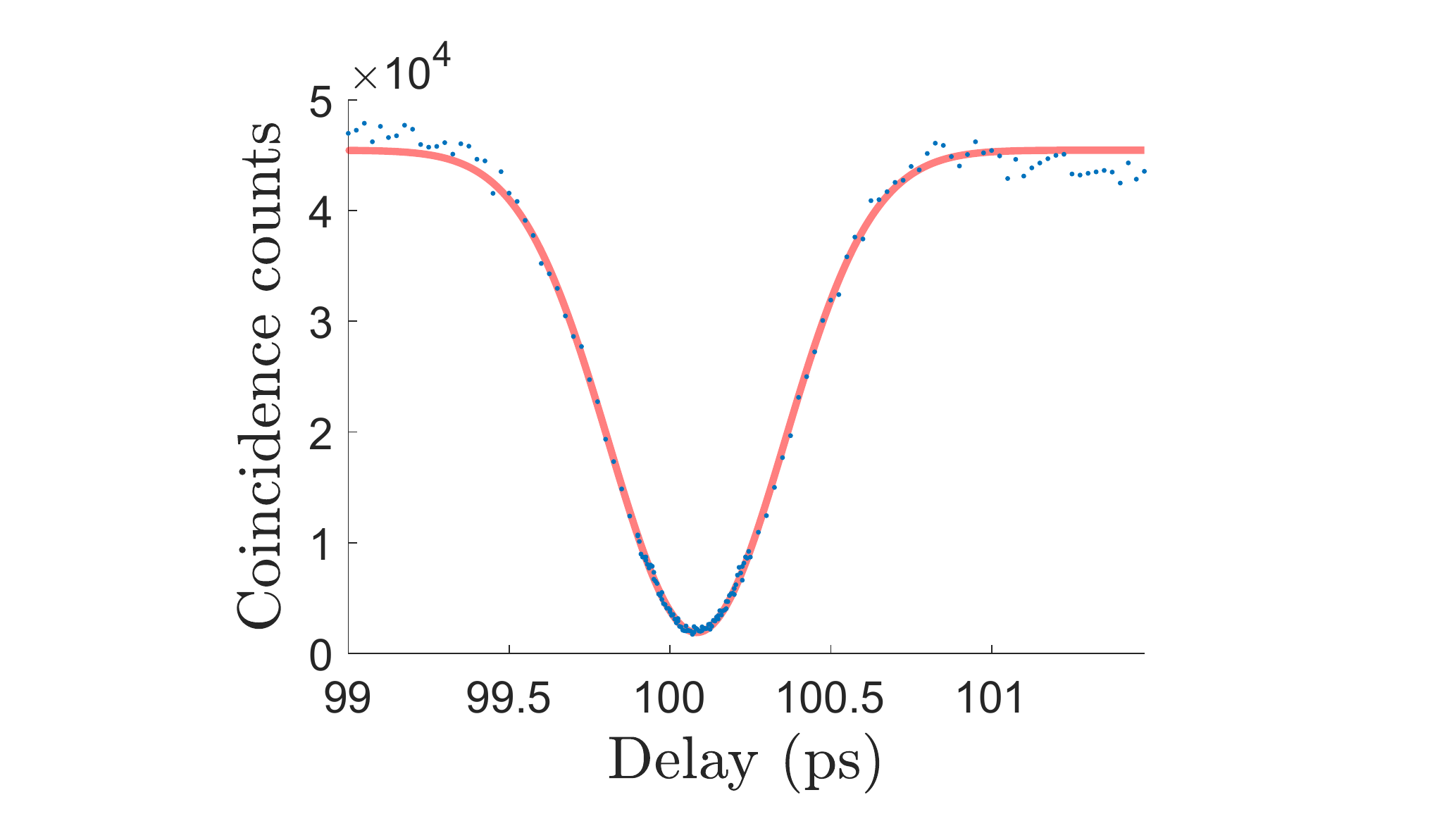}
        	\caption{\textbf{HOM interference of photon source.} Standard HOM interference measurement of the two-photon source, using a fiber-splitter to give the interference. The HOM visibility is $93\%$, without subtracting the background counts. }
        	\label{fig:1}
        \end{figure}

    \begin{figure}
    	\centering
    	\includegraphics[width = \textwidth]{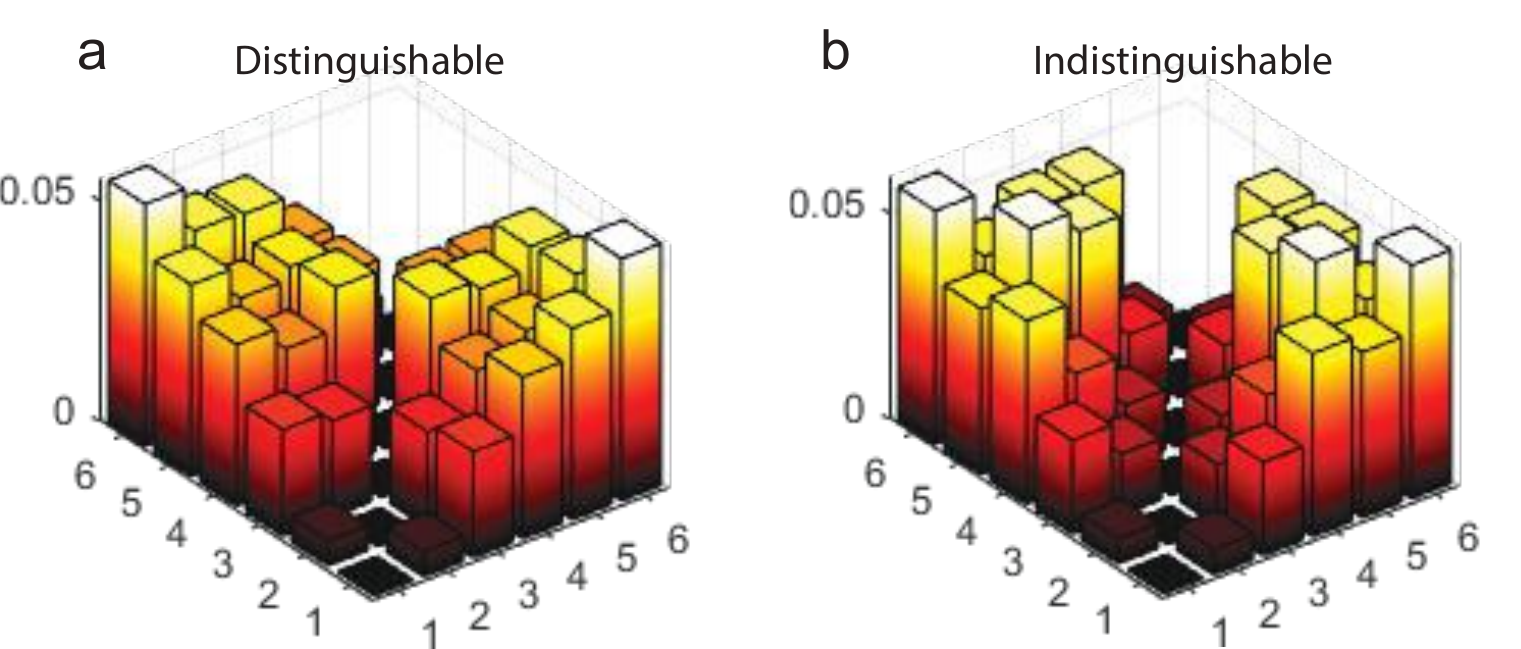}
    	\caption{\textbf{Two-photon spatial correlations of anti-bunched two-photon states at the output of the photonic chip.} (a) Correlations between the six output waveguides when the photons in the anti-bunched state are separated in time (by $> 3$ ps). 
        (b) Correlations between the six output waveguides when the photons in the anti-bunched state are overlapping in time (with $\approx 93\%$ visibility). 
    	}
    	\label{fig:2}
    \end{figure}

To demonstrate this we use a two-photon source that can be tuned between producing an anti-bunched state of distinguishable or indistinguishable photon pairs. The distinguishably is achieved by introducing variable time delay between the two photons. The quality of this source is checked using traditional HOM interference, showing a visibility of $93\%$ [Fig.~\ref{fig:1}].
    With this photon source we can therefore create two different anti-bunched two-photon states, one state where the two photons are distinguishable because of a time delay between them, and one state where there is no time delay so the photons can't be distinguished. These correspond to non-degenerate and degenerate two-photon density matrices, respectively, both of which can be described by Eq.~\ref{full_D}.

\begin{figure}
    	\centering
    	\includegraphics[width = 0.7\textwidth]{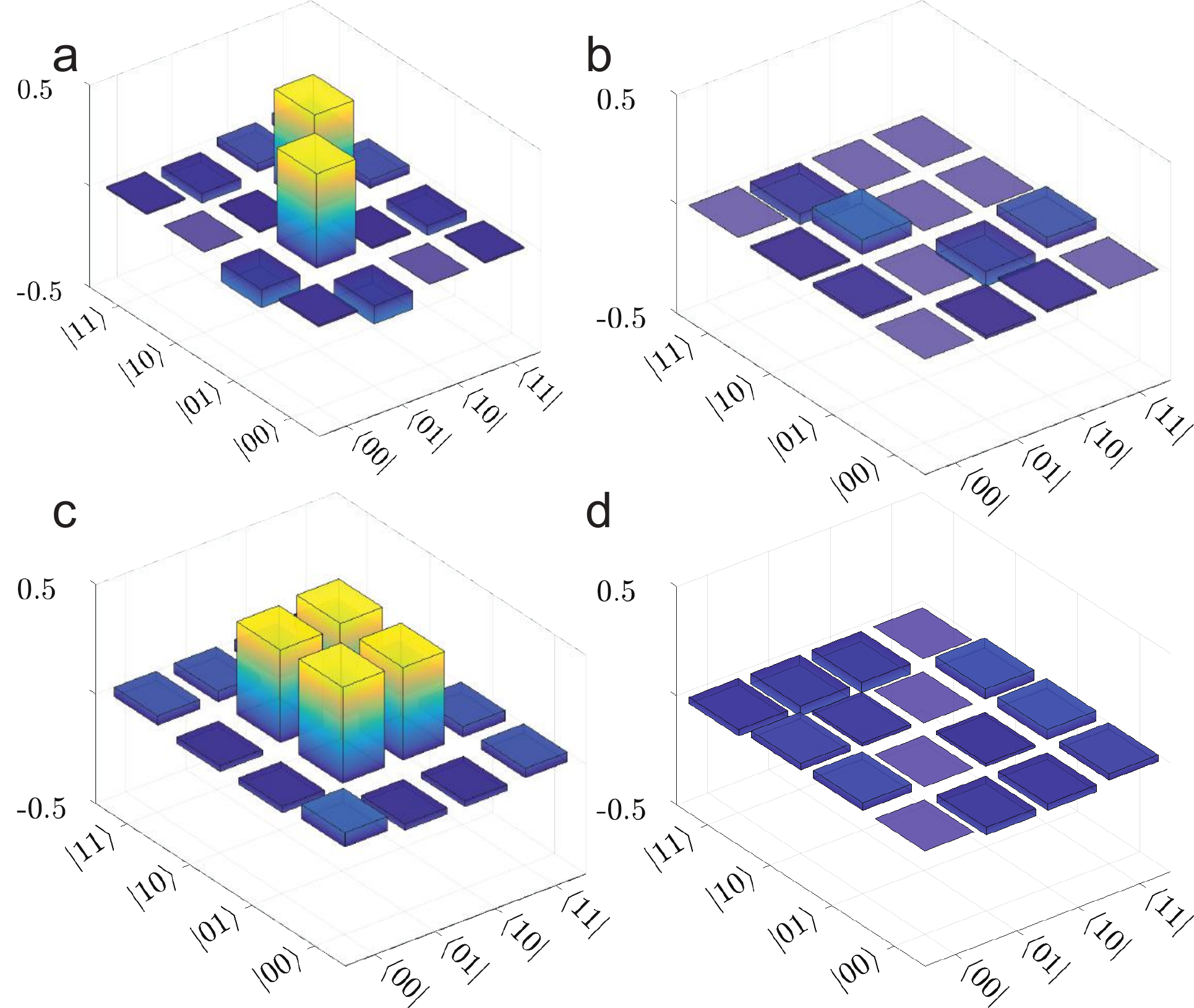}
    	\caption{\textbf{Reconstructed density matrices of distinguishable and indistinguishable photon-pairs.} (a,b) Reconstructed real and imaginary parts, respectively, of the distinguishable two-photon anti-bunched state density matrix, corresponding to the measured correlations in Fig.~\ref{fig:2}(a). (c, d) Reconstructed real and imaginary parts respectively of the indistinguishable two-photon anti-bunched state density matrix, based on the measured correlations in Fig.~\ref{fig:2}(b).
    	}
    	\label{fig:3}
    \end{figure}

    We then demonstrate recovery of both the distinguishable and indistinguishable anti-bunched states.
     We couple these two different states into our photonic chip, and observe the two distinct sets of two-photon correlations at the six outputs, much like in traditional HOM. For these two cases the two-photon correlations that are observed across the six output waveguides of the device are shown in Fig.~\ref{fig:2}(a) and  \ref{fig:2}(b) for the distinguishable and indistinguishable cases respectively. There is a clear difference between these two sets of measurements, especially in the elements $(2,3)$, $(3,4)$, and $(4,5)$, which are much larger in the distinguishable case than the indistinguishable case. This is due to quantum interference as some parts of the two-photon wavefunction are non-orthogonal in the indistinguishable case and thus interfere with one another. Thus we can observe a generalized form of the HOM effect in our device.

    As can be seen in Eq.~\ref{full_D} the form of the density matrix we are recovering depends on the overlap between the (complex) spectrum of the photons being measured, so it contains information about how indistinguishable the photons are. Performing reconstruction of the input density matrices of the distinguishable and indistinguishable coincidence measurements from Fig.~\ref{fig:2}(a) and (b) allows us to see the difference between the density matrices. The real and imaginary parts of the reconstructed density matrix for the distinguishable case is shown in Figs.~\ref{fig:3}(a) and (b) and the density matrix for the indistinguishable case is shown in Figs.~\ref{fig:3}(c) and (d). Both reflect the expected form of the reduced density matrix that was derived in Eq.~\ref{full_D}. The density matrix of the distinguishable case has small off diagonal elements, suggesting that it is a `classical' state. In contrast the density matrix of the indistinguishable state has two large off diagonal elements, these show that there is quantum coherence between different states of the two particles, and thus the state will exhibit some non-classical statistics.

    Therefore, even without photon distinguishing detection schemes,  our approach to quantum tomography is capable of recovering distinguishable and indistinguishable multi-photon quantum states using generalized HOM interference.

    %%%%%%%%%%%%%%%%%%%%%%%%%%%%%%%%%%%%%%%%%%%%%%%%%%%%%%%%%%%%%%%%%%%%%

%\author{James Titchener}
%\author{Markus Gr{\"a}fe}
%\author{Ren{\'e} Heilmann}
%\author{Alexander S. Solntsev}
%\author{Alexander Szameit}
%\author{Andrey A. Sukhorukov}

%\bibliographystyle{unsrt}
%\bibliographystyle{amsplain}
%\bibliographystyle{aps3Au4}
%\bibliographystyle{nature}

%\bibliographystyle{unsrtnat}
%\bibliographystyle{ref_links}
\bibliography{db_Scalable_onchip_tomography_v3}
%\bibliography{db_Scalable_onchip_tomography_v2,Tomography_Paper}
%\bibliography{Tomography_Paper}

\end{document}